\RequirePackage{fix-cm}
\documentclass[twocolumn,epjc3]{svjour3}

\smartqed  
\RequirePackage{graphicx}
\RequirePackage{cite} 

\RequirePackage{hyperref}
\RequirePackage{amsmath}
\RequirePackage{comment}

\journalname{...}
\begin{document}

\title{Low Energy Neutrino Detection with a Portable Water-based Liquid Scintillator Detector
}

\author{Ayse Bat\thanksref{addr1}
        \and
        Emrah Tiras\thanksref{e1, addr1, addr2} 
        \and
        Vincent Fischer\thanksref{addr3}
        \and 
        Mirac Kamislioglu\thanksref{addr4}
}

\thankstext{e1}{e-mail: etiras@fnal.gov \& emrahtiras@erciyes.edu.tr}


\institute{Department of Physics, Faculty of Science, Erciyes University, Kayseri, Turkey.\label{addr1}
            \and
           Department of Physics and Astronomy, The University of Iowa, Iowa City, Iowa, USA.\label{addr2}
            \and
           Department of Physics and Astronomy, Iowa State University, Ames, Iowa, USA.\label{addr3}
            \and
Medical Imaging Department, Vocational School of Health Services, Bandırma Onyedi Eylül University, Bandırma, Turkey.\label{addr4}
}

\date{Received: date / Accepted: date}

\maketitle

\begin{abstract}
In this study, the conceptual design and physics simulations of a near-field Water-based Liquid Scintillator (WbLS) detector placed 100 meters from the Akkuyu Nuclear Power Plant (ANPP), currently under construction and aiming at being Turkey's first nuclear power plant, is presented. 
The ANPP is an excellent opportunity for neutrino studies and the development of an R\&D program for neutrino detectors in Turkey. 
The Reactor Neutrino Experiments of Turkey (RNET) program includes a compact and portable detector with a 2.5-ton volume of WbLS and a $\sim$30\% photo-coverage, and the program is planned to be expanded with a medium-size 30-ton detector that will be an international testbed for low energy neutrino studies for WbLS and new detector technologies. 
In the following, the focus will be on the smaller $\sim$2.5~ton detector, instrumented with 8-inch high quantum efficiency PMTs and two layers of cosmic veto paddles, covering all sides of the detector, to track and veto cosmic particles. 
Inverse Beta Decay (IBD) events from electronic antineutrinos generated in the reactor core are simulated using the RAT-PAC simulation package and several liquids with different percentages of Liquid Scintillator (LS) and Gadolinium (Gd) are investigated. 


\end{abstract}

\section{Introduction}
\label{sec:intro}
While being the most abundant massive particles in the universe, neutrinos still retain many of their mysteries. Neutrino studies have come a long way since 1930 when Pauli, in his famous letter  "Dear radioactive ladies and gentlemen", proposed the neutrino as "the little one" with the words of "...I have predicted something which shall never be detected experimentally!" \cite{Pauli:1930pc}. The first experimental evidence for the existence of neutrinos came out in 1956, 26 years after Pauli postulated them, and electron antineutrinos were discovered by the Cowan-Reines neutrino experiment \cite{Cowan103}. It was followed by the discovery of muon neutrinos at Brookhaven National Lab. \cite{PhysRevLett.9.36}, the discovery of tau neutrinos \cite{KODAMA2001218} at Fermilab and finally the observation of neutrino oscillations, between the aforementioned flavors, by the SuperK experiment in Japan and SNO experiment in Canada \cite{T2K:2011qtm, T2K:2013bqz,BELLERIVE201630, SNO:2002tuh}.

The last decades changed the way we look at neutrinos, especially with its stance against the Standard Model, with the experimental discovery of the neutrino's mass. But, there are still many undiscovered facts about these mysterious particles. In order to uncover their unknown properties, the experimental techniques and technologies are evolving. 

Among these new technologies, one can find Water-based Liquid Scintillator (WbLS), Gadolinium-doped medium, and fast-timing photo-detectors such as Large area Picosecond Photo-Detectors (LAPPDs).
\begin{itemize}
    \item Water-based Liquid Scintillator: A novel detection medium that consists of a mixture of a small volume of Liquid Scintillator (LS) with a large volume of water. The high percentage of water provides both a low manufacturing cost and an easier experimental handling of the detector medium. Beyond that, this liquid emits both Cherenkov light from water and scintillation light from the LS, in a ratio set by the LS concentration. Cherenkov light provides information about the direction of charged particles in the medium with a cone-like topology. Scintillation light gives a high light yield and allows the detection of low energy charged particles below the Cherenkov threshold\cite{Kaptanoglu_2019, PhysRevD.103.052004, Gab2020, onken2020time, Gab2021}.
    \item Gadolinium doped medium: When it comes to detecting neutrons in hydrogenated media, Gd and its high thermal neutron capture cross-section (49,000 barn), compared to Hydrogen (0.3 barn), is an element often chosen by physicists. Gd has a moderate price and is easy to incorporate in aqueous solutions thanks to its highly soluble nature in water. \cite{PhysRevLett.93.171101, vagins2007gadzooks, DAZELEY2009616, back2020measurement}
    \item Large Area Picosecond Photo-Detectors (LAPPDs): A new photodetector technology, based on the principle of micro-channel plates, capable of reaching time resolutions of a few tenths of pico-seconds and high spatial resolutions down to the mm scale. The LAPDDs are significantly promising on separation of prompt Cherenkov light from delayed scintillation light\cite{LAPPD:2016yng,LYASHENKO2020162834,adams2013measurements, minot2018large, tiras2019detector}.
\end{itemize}

Neutrinos come from different natural and man-made sources such as cosmic rays, supernovae, atmospheric muon decays, the Sun, accelerators and nuclear reactors. The first experimental evidence of neutrinos came from a nuclear reactor in 1956 \cite{Cowan103} and since then, reactor neutrinos have been studied extensively. As they are intensely generated in only one flavor, $\bar{\nu_{e}}$, reactor neutrinos have been particularly helpful for neutrino oscillation studies where the flavor change is observed through a disappearance of the initial antineutrino flux. 

Nuclear power is at the forefront of energy policies as it provides opportunities such as obtaining energy at an affordable price while meeting energy demands, developing nuclear technology for peaceful applications, providing carbon-free energy production and supporting the development of alternative energy sources. 
Conducting scientific studies, such as neutrino physics, in countries that are developing this technology for the first time could provide additional opportunities for training people-power in the field of physics and nuclear engineering.

The emission of electron anti-neutrinos in a reactor core is directly correlated with the fission rate and the thermal power of the core and the emission rate, $N_\nu$, can be expressed as:
\begin{equation}
N_\nu=\gamma(1+k)P_{th}
\end{equation}
where $\gamma$ refers to a constant number containing all invariable terms, $P_{\text{th}}$ indicates the reactor thermal power and $k$ represents a time-dependent factor representing the variation of fuel composition over time. 
For the remainder of this study, this $k$ factor will be assumed equal to zero and time-dependent fuel composition will be considered constant.
Assuming the reactor thermal power is constant over time, a safe assumption for commercial reactors focusing on power generation, the rate of anti-neutrinos emission is considered constant over time.

About $200$~MeV of energy is emitted per fission and most of this energy is shared between multiple fission products.
These neutron-rich nuclei then undergo a series of beta decays leading to the total average emission 6~$\bar{\nu_{e}}$ per fission.
The total amount of neutrinos emitted as a function of the reactor thermal power is approximately $2 \times 10^{20}$ $\bar{\nu_{e}}$ GW$_{th}^{-1}$.s$^{-1}$.
The energy distribution of those neutrinos, centered around $\sim$3~MeV is mostly driven by the isotopes at the origin of the fission, $^{235}U$, $^{238}U$, $^{239}Pu$, or $^{241}Pu$ \cite {HICKEY2021101961, PhysRevD.53.6054}.

In uranium-based nuclear reactors, as the fuel composition develops, the initial uranium content gradually decreases while the plutonium content increases. 
This effect, known as burn-up, changes the number of neutrinos released and their energy distribution over time. 
For this study, this energy distribution, as well as the neutrino emission rate, will be considered constant \cite {kuleff1984neutron}. 

The Akkuyu Nuclear Power Plant (ANPP), Turkey's first nuclear power plant, is planned to start operating in 2023 in Mersin. 
It is the product of a cooperation agreement signed between the governments of Russia and Turkey in 2010 and its construction has started in 2018. 
The power plant will be equipped with four AES-2006 Generation III+ VVER units, a new generation pressurized water reactor constructed by the Rosatom company, fueled with uranium dioxide enriched to 5\% $^{235}U$. 
The nominal electrical power of each reactor will be approximately 1200~MW$_{\text{e}}$, for a thermal power of 3200~MW$_{\text{th}}$.
Upon the successful operation of the first reactor, Akkuyu-1, the 2$^{\text{nd}}$, 3$^{\text{rd}}$, and 4$^{\text{th}}$ units will be put into operation in 2024, 2025, and 2026, respectively \cite{hayes2016reactor, ulgen2015turkey}. 

The rate of anti-neutrino interactions expected in a detector can be expressed as:
\begin{equation}
S_\nu=\frac {N_{f}N_{p}}{4{\pi}L^2}<\sigma>
\end{equation}
where, $N_p$ is the number of protons in the detector and $L$ indicates the distance from the reactor core to the neutrino detector. The $<{\sigma}>=5.8 x 10^{-43}$ is the average inverse beta decay cross section at reactor neutrino energies and $N_f$ represents the average fission rate, which is obtained from the following equation \cite{anjos2015using}: 
\begin{equation}
N_{f}=6.24 \times 10^{18} (\frac{P_{th}}{MW^{-1}})(\frac{MeV}{W_{e}})s^{-1}
\end{equation}
where the mean energy delivered per fission is given by $W_{e} = 203.78MeV$. If we calculate the neutrino interaction rate for the Akkuyu NPP, we obtain:
\begin{equation}
S_\nu=9.86 \times 10^{5}(\frac{V}{m^3})(\frac{m}{L})^2 \quad \text{events/day}. 
\end{equation}

The anti-neutrino interaction rate (in events/day) for different reactor-detector distances is shown in Table 1.
In this table, $V$ is the detector volume and $L$ represents the distance between the reactor core and the detector \cite{cao2012determining}. 
Detailed information regarding the detector geometry and size can be found in Section~\ref{sec:2}.

\begin{table}[!bh]
\centering
\caption{The number of reactor neutrinos per day (events/day) for neutrino interactions at various reactor-detector distances.}
\label{tab:trigo}
\begin{tabular}{|c c c|}
\hline
V($m^3$)         & L(m)    & $S_v$ (events/day)\\
\hline
2,65        & 100          & 261              \\
\hline
2,65        & 50           & 1050              \\
\hline
2,65        & 25           & 4180  \\        
\hline
\end{tabular}
\end{table}

\begin{figure}[ht!]
    \centering
    \includegraphics[width=0.4\textwidth]{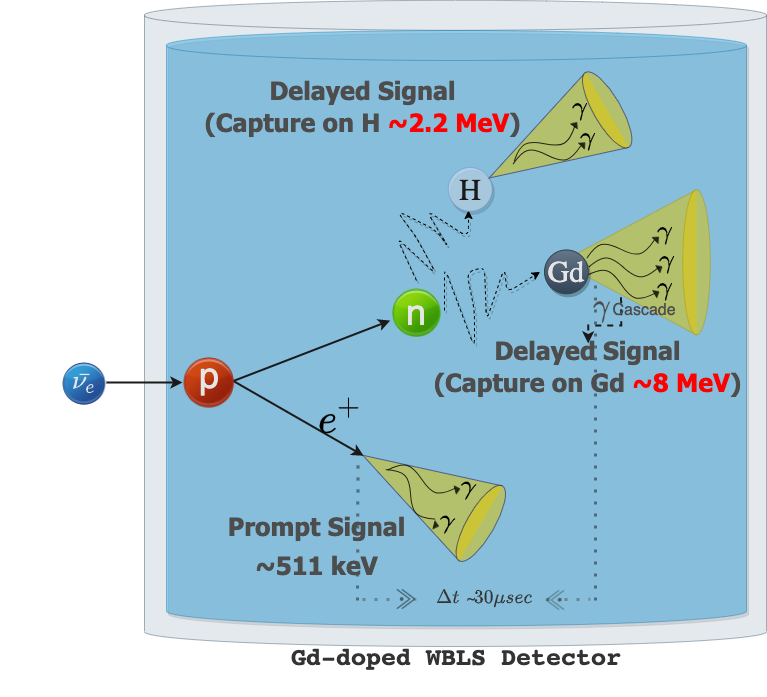}
    \caption{Inverse Beta Decay (IBD) process and signal reconstruction in the Gd-doped Water-based Liquid Scintillator (WbLS) detector.}
    \label{fig:IBD}
\end{figure}

Electronic anti-neutrinos,  produced in the beta decays of fission products in nuclear reactors, are detected via Inverse Beta Decay, or IBD ($\bar{\nu} + p \rightarrow e^+ + n $). 
This reaction leads to the creation of two distinct signals, one from the positron, the other from the neutron, both separated in time and with different signatures. 
The first signal comes from the annihilation of the positron with an electron in the detection medium and leads to the emission of two 511~keV gamma rays. This process is referred to as a prompt signal, $E_{prompt}$.  
The second signal, generated by the capture of the neutron on an atom in the medium, occurs several microseconds ($\sim 10-100\mu \text{sec}$ depending on the detection medium) after the prompt signal. 
Most IBD detectors utilize hydrogenated materials (water, liquid scintillator) loaded with isotopes having a high neutron capture cross section in order to efficiently and quickly thermalize and capture neutrons. 
For most of the aforementioned detectors and for the rest of this study, Gadolinium is the isotope being used. 
Radiative neutron capture on H releases $\sim 2.2$ MeV of energy in the form of a single gamma ray, while neutron capture on Gd releases a gamma cascade with a total energy of about $8$ MeV. These captures lead to a delayed signal, $E_{delayed}$, with a visible energy higher than most backgrounds due to natural radioactivity.  
A sketch of the IBD detection in a detector is shown in Fig.\ref{fig:IBD}.

\section{Low Energy Neutrino Studies in Turkey}
\label{sec:1}
In recent years, several design and simulation studies have been performed with the goal of detecting neutrinos from the Akkuyu NPP.
The first study conducted by S. Ozturk et al. is a Geant-4 based simulation of a small, 1-ton Gadolinium-doped water Cherenkov detector in very close vicinity from one of the reactor cores \cite{Ozturk:2016tjn}. 
Another study, by S. Ozturk, focused on using a segmented detector, made of Gadolinium loaded plastic scintillators. 
In this study, he showed the benefits of using a multi-variable analysis technique to suppress cosmic backgrounds \cite{OZTURK2020163314}. 
On the other hand, M. Kandemir and A. Cakir proposed a small segmented hexagonal detector composed of plastic scintillators attached to photomultiplier tubes (PMTs) \cite{KANDEMIR2020}. 
The common purpose of these studies is to monitor the reactor anti-neutrinos flux for nuclear non-proliferation purposes. 
But in a broader perspective, we propose to develop a program for low-energy neutrino studies in Turkey, which includes both a small size detector at a short-distance to the reactor, as mentioned here, and a medium size detector located 1-2~km from the reactor core. 
The medium size detector, previously proposed by our co-authors, V. Fischer and E. Tiras, includes a 30-ton cylindrical tank filled with 0.1\% Gadolinium doped Water-based Liquid Scintillator (WbLS) and aims to be placed a few km away from the reactor cores \cite{FISCHER2020163931}.
The Reactor Neutrino Experiments of Turkey (RNET) program will consist of both a small size detector (2-3 ton) placed very close to one of the reactor cores (about 50 m) and a far detector about 1-2 km from the power plant for low-energy neutrino oscillation and nuclear non-proliferation studies using new techniques and pioneering technologies. 
Both detectors will be complimentary and they will have versatile designs to be able to test new detectors, sub-detectors and detection media in-situ. 
They will be instrumented with same type of High Quantum Efficiency (HQE) PMTs, and the detector medium's concentration will be similar in order to help cancel systematic effects related to the detector response. 
The RNET program will be a great opportunity for neutrino studies in Turkey with the aims of; (1) monitoring the Akkuyu NPP for nuclear non-proliferation purposes \cite{BOWDEN2007985, bernstein2008monitoring, ashenfelter2016prospect, haghighat2020observation}, (2) being a testbed for new detector technologies and new media as an integration with neutrino physics collaborations around the globe, (3) researching and developing more sensitive detectors and (4) training the next generation of neutrino physics researchers.

\section{Experimental Setup}
\label{sec:2}
The experimental setup, designed to be portable, includes a cylindrical tank, two layers of cosmic veto paddles on the sides of the tank, an electronics rack, and a DAQ system. 

\begin{figure}[!t]
    \centering
    \includegraphics[width=0.4\textwidth]{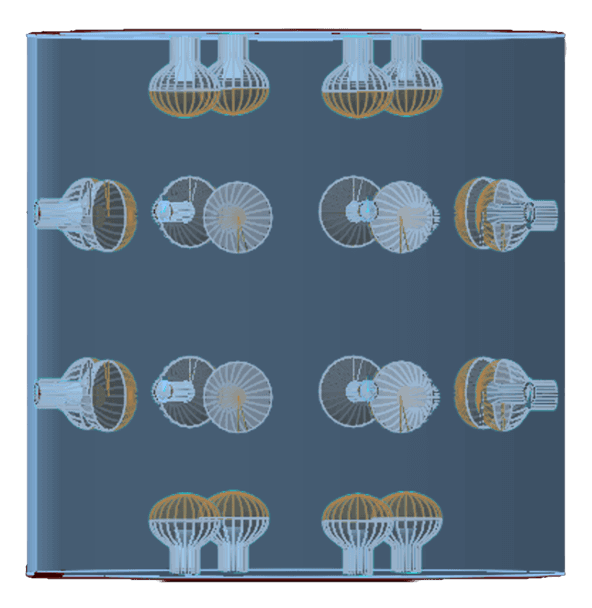}
    \caption{Visualization of the detector and its inner PMTs using the RAT-PAC simulation framework.}
    \label{fig:ratpac}
\end{figure}

\begin{figure}[!h]
    \centering
    \includegraphics[width=0.4\textwidth]{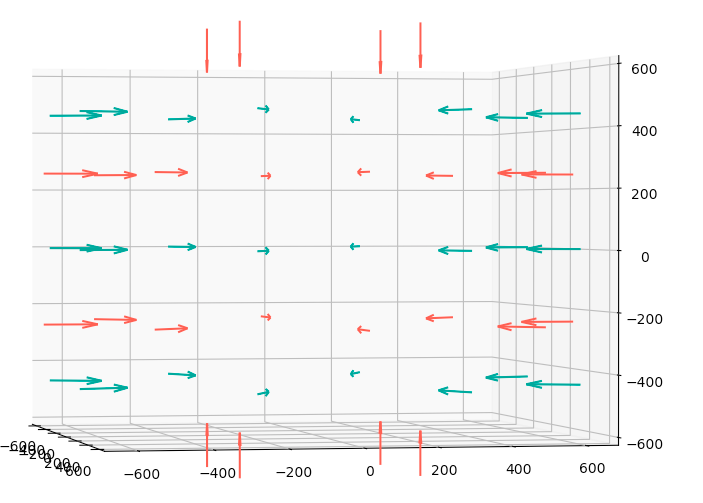}
    \caption{Sketch of the orientation of each PMTs in the detector. Each arrow indicates a PMT location in the detector and the direction of the arrow indicates the direction of the face (photocathode) of the PMTs. Orange arrows corresponds to 8-inch PMTs while green arrows correspond to possible additional 2-inch PMTs.}
    \label{fig:2inch}
\end{figure}

The tank, a stainless steel cylinder 1.5-m tall and 1.5-m in diameter, will be filled with a water-based liquid scintillator solution with a 0.1\% Gd concentration and a 3\% liquid scintillation concentration.  
The detector will be instrumented with 24 8-inch HQE PMTs facing the inside the detector and placed homogeneously around the detector barrel and on the top and bottom of the detector.
Such a PMT placement corresponds to a photo-coverage of 28\%. 
Fig.~\ref{fig:ratpac} shows the RAT-PAC generated visualization of the detector, including the locations of all its PMTs in the inner volume. 
Several 2-inch PMTs could also be placed in the gaps between 8-inch PMTs to increase photo-coverage and provide additional granularity for light pattern reconstruction.
Fig. \ref{fig:2inch} shows the location and direction of all types of PMTs in the detector.
For the remainder of this study, only the 8-inch PMTs will be taken into account. 

The cosmic ray veto will consist of layers of plastic scintillator paddles attached to light guides ($\sim 20$-cm long) and coupled with 2-inch PMTs. 
Each layer, made of several scintillator paddles, is expected to have dimensions of at least 1.7 m (width) x 1.7 m (length) in order to efficiently cover all sides of the detector for cosmic ray tagging purposes. 
Plastic scintillator was chosen as the main detection material for this veto as it has a high light yield and a fast scintillation decay time, thus making them a effective and inexpensive choice for efficiently rejecting cosmic backgrounds.

Inside the tank, a PMT holding structure, machined out of stainless steel attached to the tank lid, will be holding all PMTs.
This design allows for an easy removal of each photosensors in case of repairs or upgrades.

Once the PMTs are mounted on the structure, the inner structure will be covered with a white reflective and gadolinium-compatible plastic tarp, in order to increase light collection as well as isolate the inner sensitive volume from neutron captures occurring in the outer volume of the tank. 

The lid and tank cylinder will be equipped with handles in order to easily lift and move the detector using a simple winch mechanism.
With the added benefits of allowing quick and easy maintenance operations on the detector, this will also make the detector more portable and able to be deployed at various locations.

\section{Electronics and Data Acquisition System}
\label{sec:3}
The whole detector system will be based on three major components: a Gd-doped WbLS tank, a cosmic muon veto, and a trigger and Data Acquisition System (DAQ) system.

The trigger system will rely on the coincidence-based detection of light depositions in the tank.
Signals from the water PMTs will be received by VME-based 500~Mhz analog to digital converters (ADC) used for waveform digitization. 
The signals from PMTs part of the cosmic veto system will be read out through a NIM/CAMAC system and fed into Time-to-Digital Converters (TDC) that will only the hit times of each paddles and allow a simple reconstruction of the muon track in the detector.

\begin{figure}[!t]
    \centering
    \includegraphics[width=0.4\textwidth]{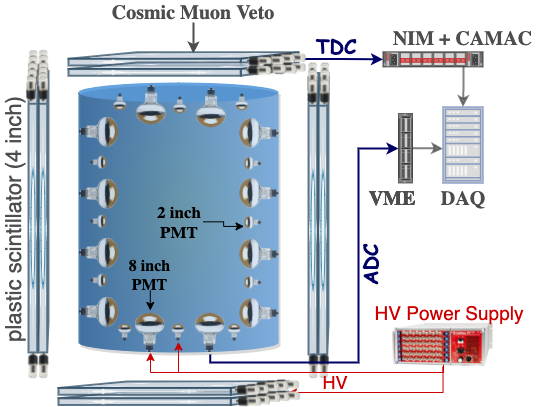}
    \caption{Schematic view of the detector components, electronics and DAQ system.}
    \label{fig:detector}
\end{figure}

The whole system will be powered by a high voltage power supply, capable of providing positive (tank PMTs) and negative (veto PMTs) polarities. 
Fig.~\ref{fig:detector} shows the schematic view of the detector components, electronics and Data Acquisition (DAQ) system.

\section{Simulation Studies}
\label{sec:4}
\subsection{Signal Simulation}
In order to simulate particles transport in an accurate model of the Gd-doped WbLS detector, we used RAT-PAC (Reactor Analysis Tool - Plus Additional Code), a simulation framework that utilizes Geant4 and GLG4Sim libraries.

A series of simulations have been performed to understand the detector response to IBD interactions. 
For that purpose, a build-in RAT-PAC algorithm generating realistic IBD positron-neutron pairs was used.
The kinematics (energy and momentum) of each IBD pair are obtained from the inverse beta decay differential cross section assuming a realistic ratio of fission products ($^{235}U$, $^{238}U$, $^{239}P$, $^{241}P$) in the reactor core.
For this study, this ratio of fission products is considered uniform in time.

In order to make statistical effects negligible, a total of 500,000~IBD events was simulated uniformly throughout the detector.
Electrons with a fixed energy, ranging from 1 to 10 MeV, were also simulated uniformly in the detector to determine the conversion factor between the number of photon hits, expressed after correction for the quantum efficiency as a total number of photo-electrons, and visible energy in MeV.
This conversion factor, that varies for different media, was used to display all energy distributions in terms of visible energy and better compare different detector designs. 

\begin{figure}[!bh]
     \centering
     \includegraphics[width=0.48\textwidth]{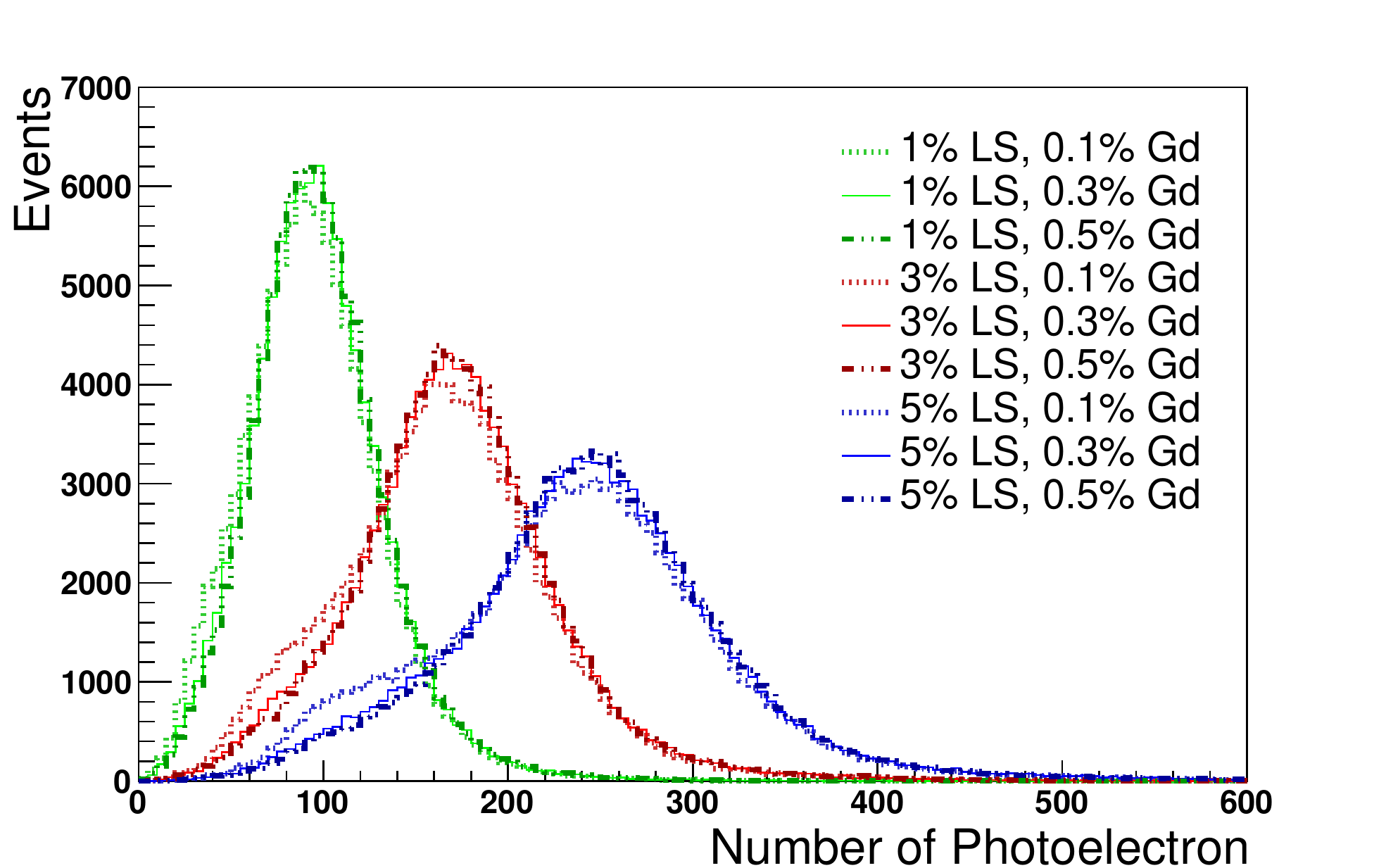}
     \caption{Photo-electron distributions of $E_{prompt}$ for corresponding WbLS cocktail with different WbLS concentrations (1\%, 3\%, and 5\% LS) and different mass fractions of Gd (0.1\%, 0.3\%, and 0.5\%).}
     \label{fig:numPE_wbls}
 \end{figure}

\begin{figure}[!bh]
     \centering
     \includegraphics[width=0.48\textwidth]{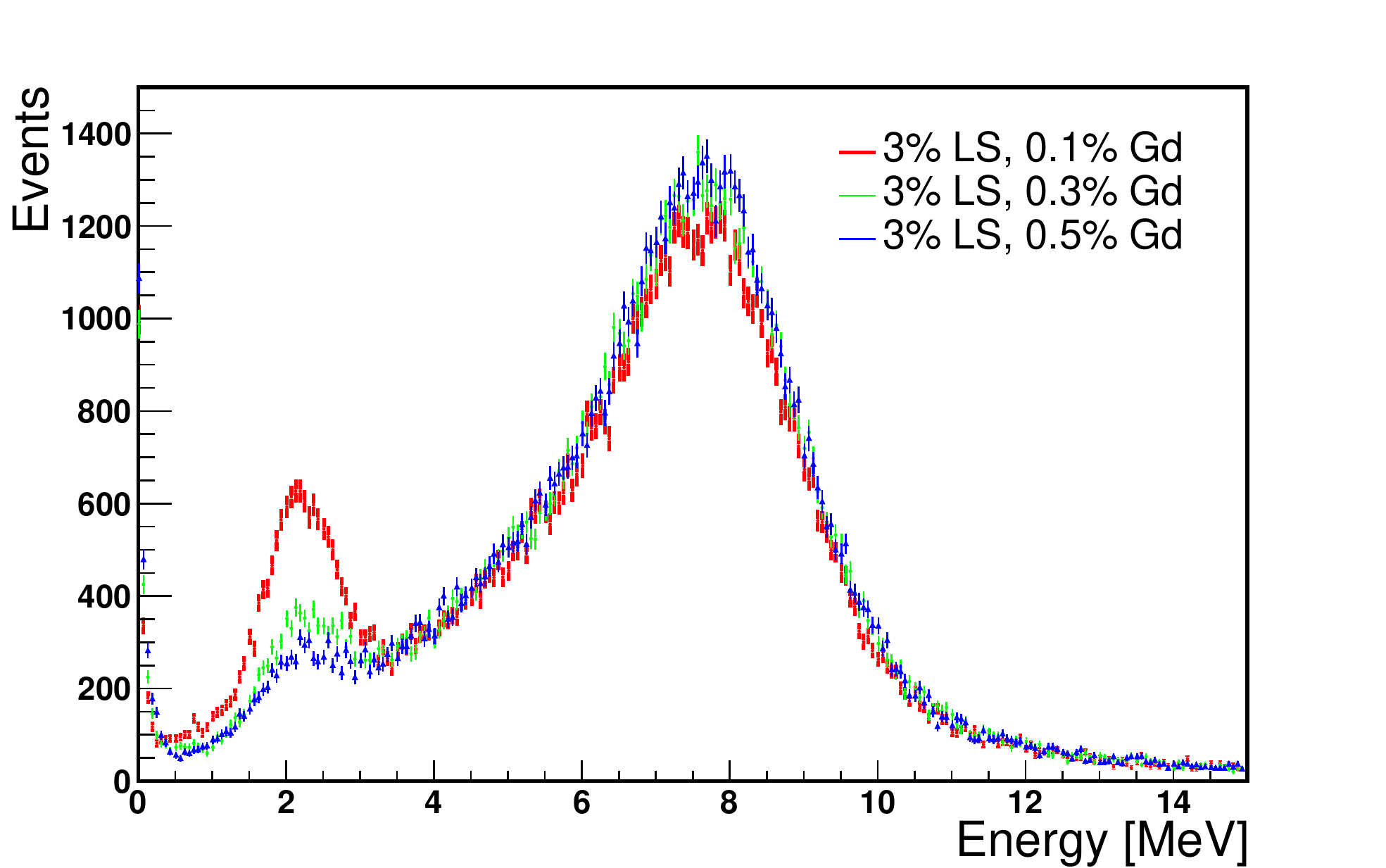}
     \caption{Reconstructed delayed energy distributions for the corresponding WbLS cocktail with different WbLS concentrations (1\%, 3\%, and 5\% LS) and different mass fractions of Gd (0.1\%, 0.3\%, and 0.5\%).}
     \label{fig:3LS_01Gd}
 \end{figure}

Several IBD events simulations were carried out with different WbLS concentrations (1\%, 3\%, and 5\% LS) and with different mass fractions of Gd. 
Fig.~\ref{fig:numPE_wbls} shows the photo-electrons distributions using different WbLS concentrations. 
As expected, the number of photo-electrons is directly correlated with the concentration of liquid scintillator in the medium, thus allowing a better efficiency at detecting low energy particles, a crucial feature when observing reactor anti-neutrinos, whose energies peak at around 2-3 MeV.
Fig.~\ref{fig:3LS_01Gd} shows the reconstructed delayed energy distributions for different mass fractions of Gd in a 3\% WbLS medium. 
Increasing the gadolinium concentration results in an enhanced efficiency to detect an n-Gd capture at a fixed threshold of 4~MeV. 

While many different WbLS solutions have already been manufactured on a small scale, the WbLS cocktail with 3\% LS and 0.1\%~Gd is one of the most studied and is expected to be manufactured on a bigger scale ($\geq$~100~tons).
This WbLS mixture is thus the one we decided to focus on for the rest of this study.
Similar simulations were also performed with a 5\% LS solution as a cross-check since it has been manufactured and thoroughly studied as well.

In the following, we discuss the energy and vertex resolutions for a 3\% LS and 0.1\%~Gd WbLS solution in our detector model. 

\begin{figure}
    \centering
    \includegraphics[width=0.48\textwidth]{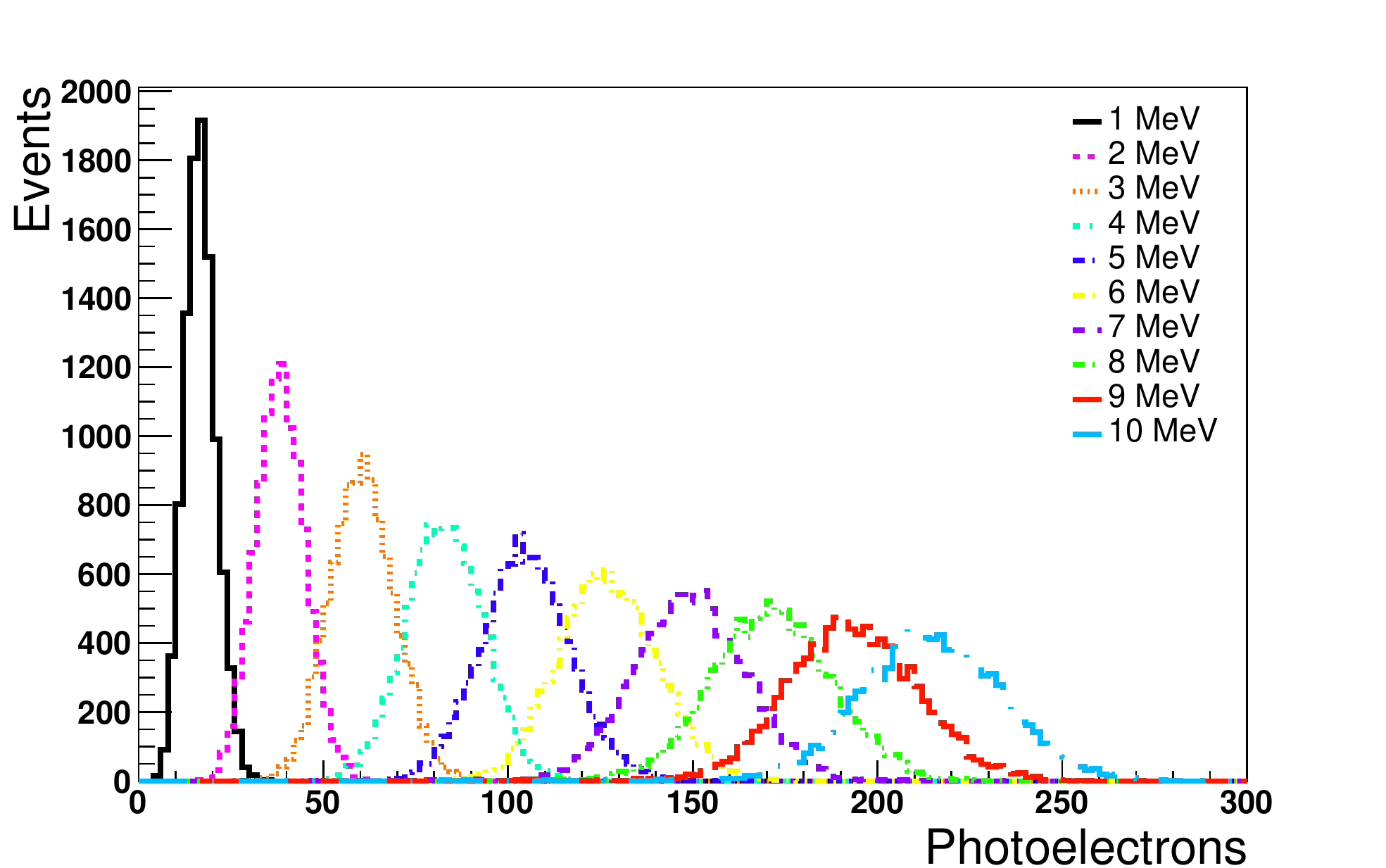}
    \caption{Photo-electron distribution as a function of energy ranging from 1 to 10 MeV with 1MeV step obtained using simulated electrons generated at the center of the detector.}
    \label{fig:pe}
\end{figure}

\begin{figure}
    \centering
    \includegraphics[width=0.48\textwidth]{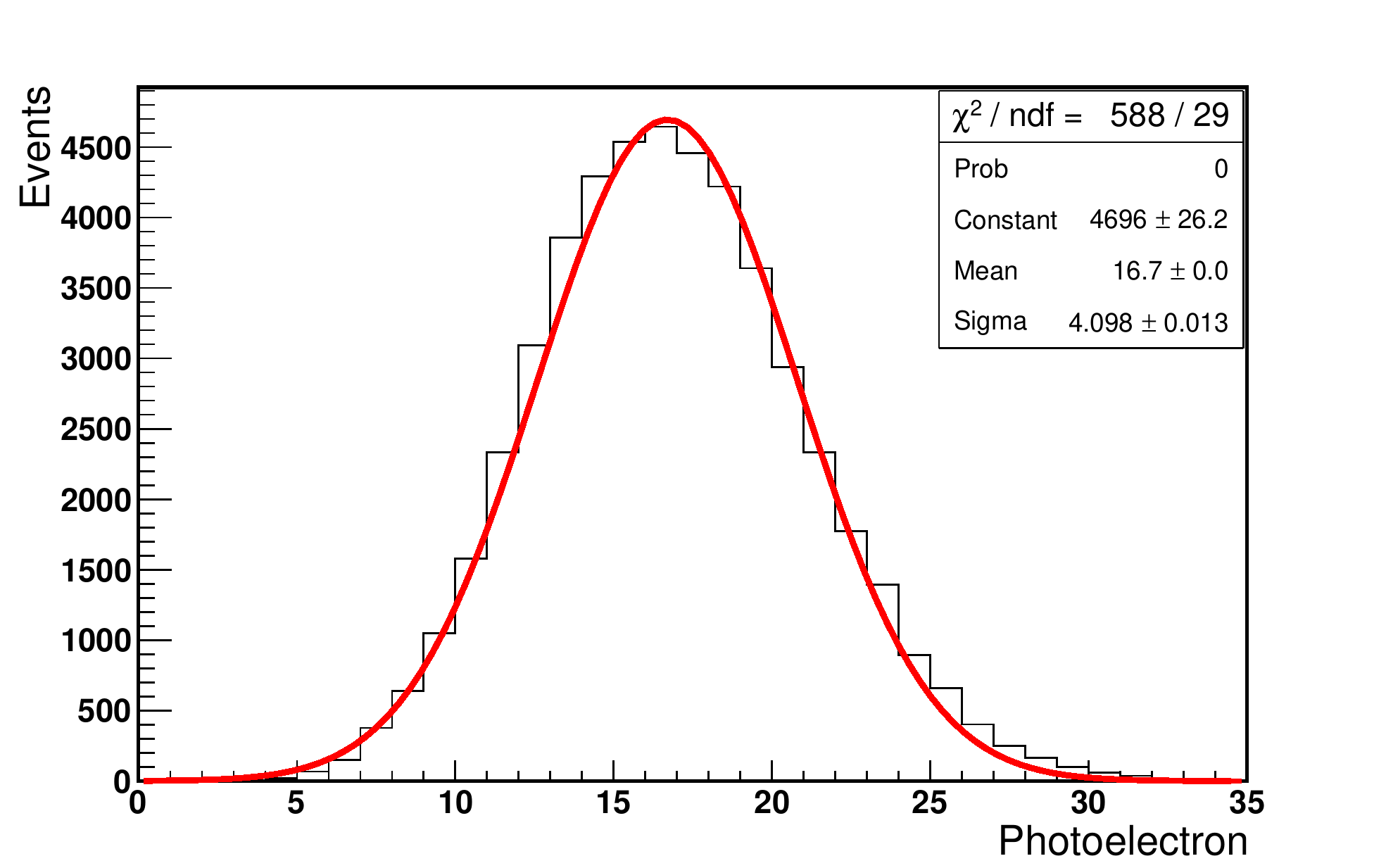}
    \caption{Photo-electron distribution of 1 MeV electrons generated at the center of the detector and fitted with a Gaussian function.}
    \label{fig:fit_MeV}
\end{figure}

A widely accepted approximation is that pure LAB-based liquid scintillators have a light yield of about 10,000 photons per MeV.
Following this approximation as well as linearly scaling the light yield with the LS concentration, an approximation valid at first order, the aforementioned medium with 3\% LS yields about 300 photons per MeV.
Fig.~\ref{fig:pe} shows the photo-electron distribution as a function of energy from simulated electrons at the center of the detector.
To obtain the energy resolution, electrons with a fixed energy of 1~MeV were simulated at the center of the detector. 
The photo-electron distribution of this electrons along with its best gaussian fit applied to it are showed in Fig.~\ref{fig:fit_MeV}.
With an average of 17 detected photo-electrons per 1~MeV electron, the gaussian fit yields an energy resolution of 24\% at 1 MeV.
Since the energy resolution directly depends on the photon statistics, the energy resolution can be extrapolated at different energies using the formula $24\%/ \sqrt{\text{E (MeV)}}$.

 
Accurately reconstructing the positions of events in a detector is essential when trying to select IBD pairs amidst backgrounds as prompt and delayed events are correlated in time and space.

\begin{figure}
    \centering
    \includegraphics[width=0.48\textwidth]{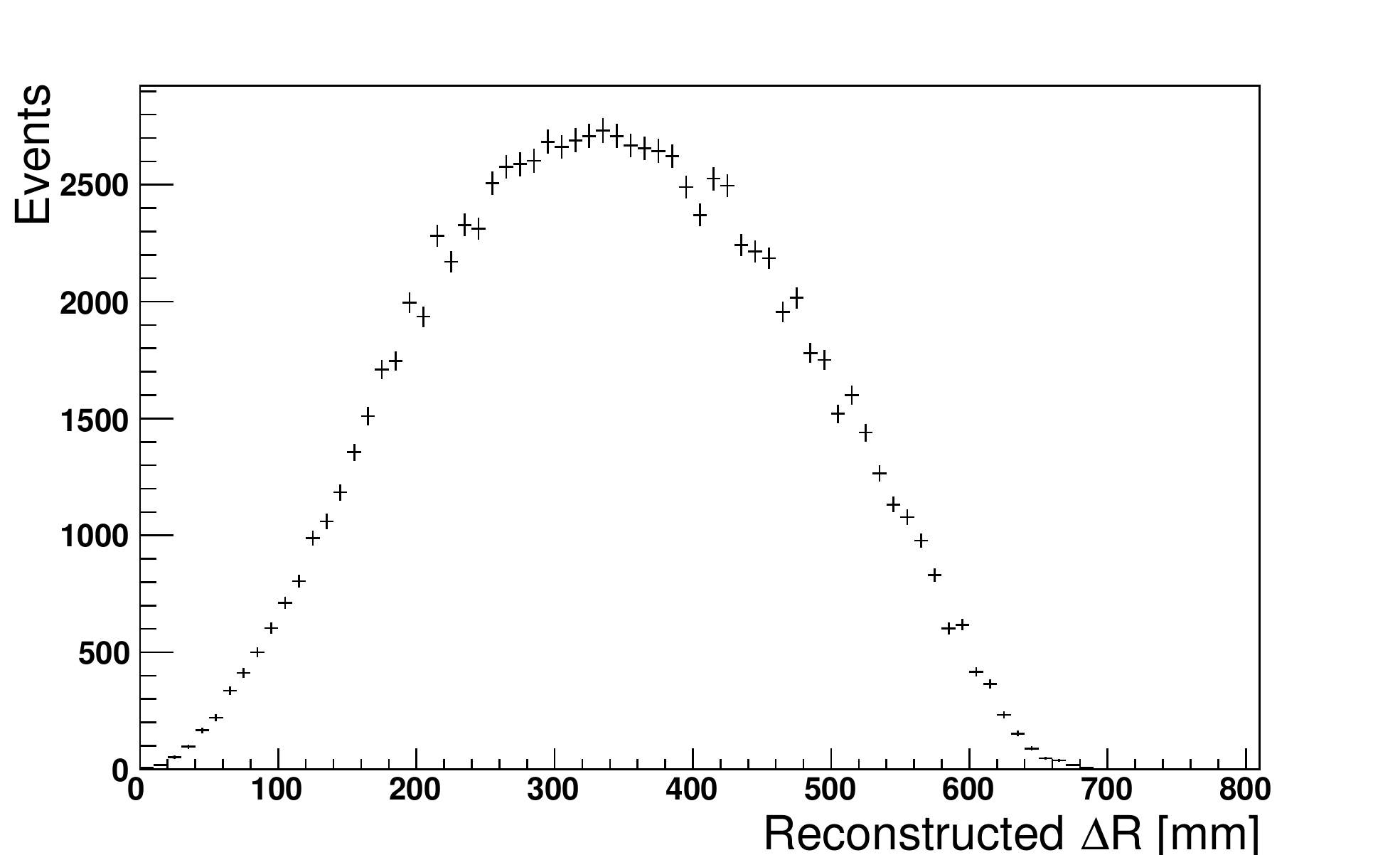}
    \caption{Reconstructed $\Delta R$ distribution from simulated 1~MeV electrons using the BONSAI energy and position reconstruction algorithm.}
    \label{fig:deltaR}
\end{figure}

\begin{figure}
    \centering
    \includegraphics[width=0.48\textwidth]{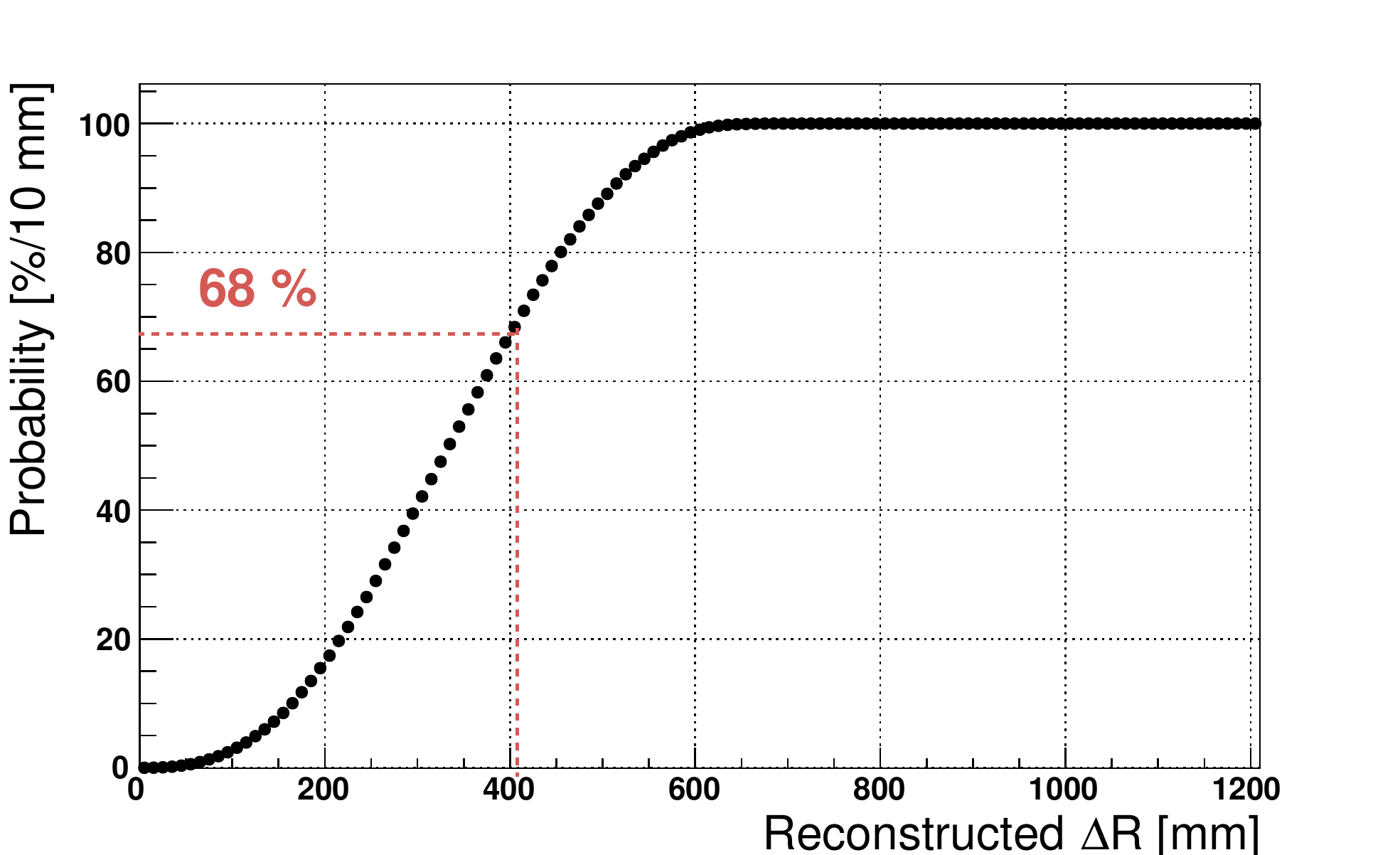}
    \caption{Cumulative $\Delta R$ distribution obtained from the reconstructed $\Delta R$ distribution displayed in Fig.~\ref{fig:deltaR}}
    \label{fig:deltaR_cumul}
\end{figure}

In the RAT-PAC simulation framework, BONSAI~\cite{Smy:2007maa} can be used as a position reconstruction algorithm to obtain the best reconstructed position and direction of a particle in the detector. 
Since BONSAI was mostly designed to study low energy events, 1~MeV electrons were simulated to verify the position reconstruction capabilities of the detector. 
The spatial separation between simulated and reconstructed vertices, shown as the distribution of $\Delta R$, is displayed in Fig.~\ref{fig:deltaR}. 
Fig.~\ref{fig:deltaR_cumul} shows the cumulative distribution of $\Delta R$ and indicates that 68\% of the interaction vertices are reconstructed within 41 cm of their true position. 
Similarly to the energy resolution, the vertex resolution is dependent of the photon statistics at first order and can thus be expressed as a function of energy through the following formula: $41 \text{cm} / \sqrt{\text{E (MeV)}}$.
Given the relatively small size of the detector, such a position resolution does not yield a significant advantage for background discrimination.
However, several studies are ongoing to improve reconstruction algorithms in WbLS and extract information from both the Cherenkov and scintillation components of the emitted light.
 

At this point, the behavior of ionizing particles in our detector model is understood enough to efficiently simulate and interpret events from IBD pairs generated by reactor neutrinos.

While positrons and neutrons are generated at the same time in the simulation, the separation between prompt and delayed events is performed in the analysis in a fashion similar to what a simple DAQ would do.
All hits (photo-electrons) on the detector's PMTs within the [0-200]~ns time range are considered part of the \textit{prompt event}, caused by the interaction of the positron in the medium.
Similarly, all hits recorded in the [1-200]~$\mu$s time range are considered to be part of the delayed event, caused by the neutron capture on gadolinium or hydrogen.

\begin{figure}
    \centering
    \includegraphics[width=0.48\textwidth]{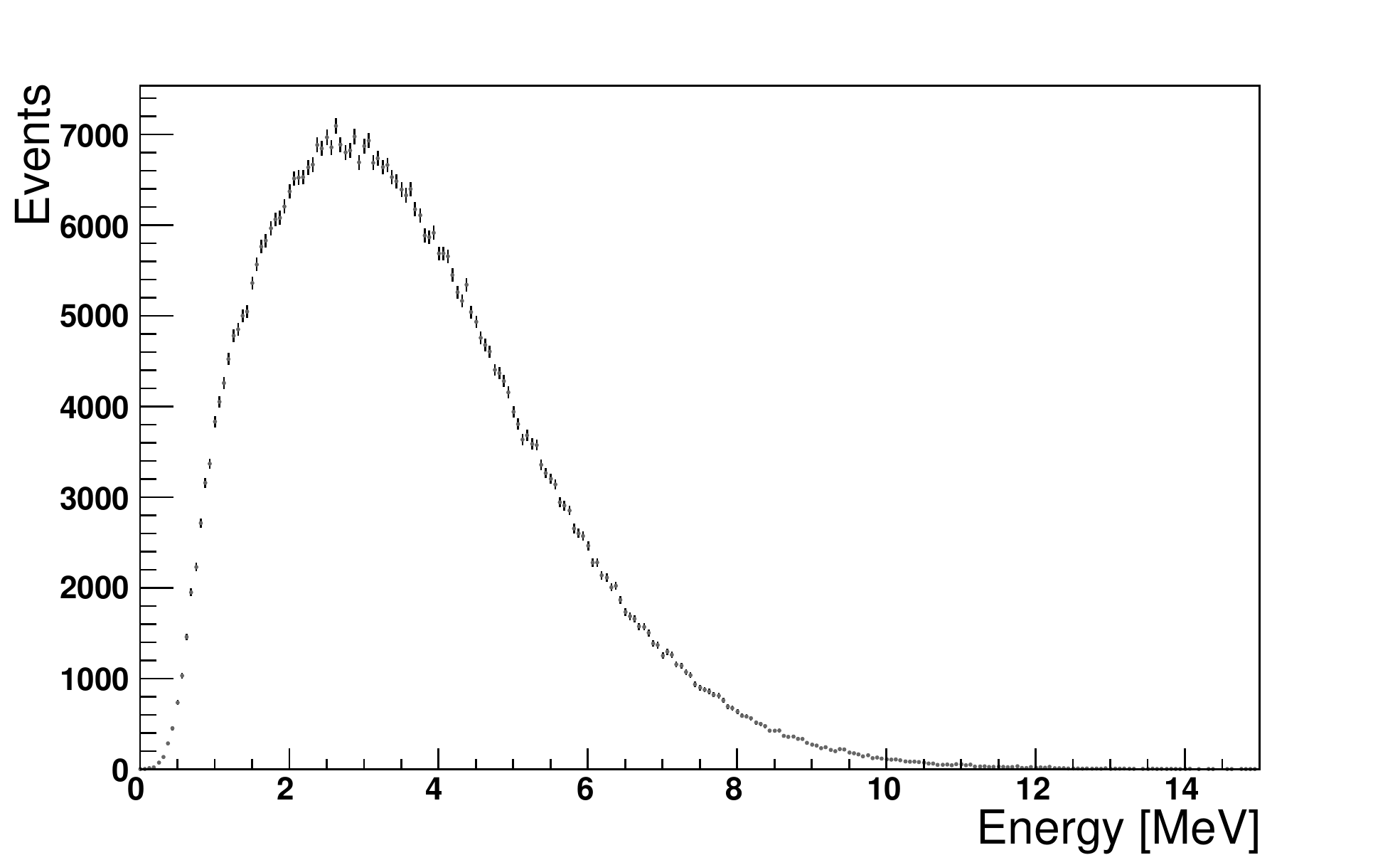}
    \caption{Reconstructed prompt events energy distributions from the simulated IBD events.}
    \label{fig:prompt}
\end{figure}

\begin{figure}
    \centering
    \includegraphics[width=0.48\textwidth]{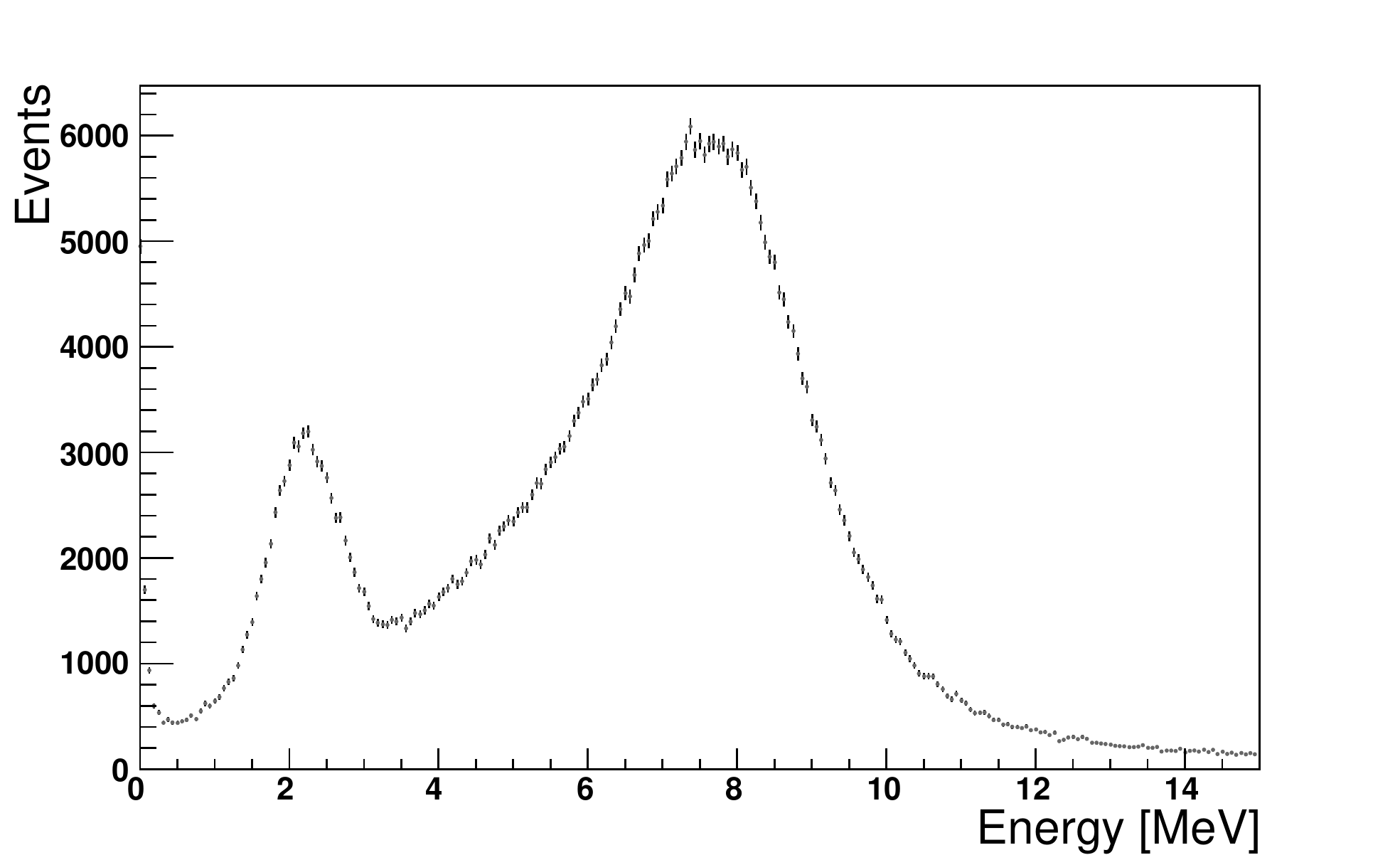}
    \caption{Reconstructed delayed events energy distributions from the simulated IBD events.}
    \label{fig:delayed}
\end{figure}

\begin{figure}
    \centering
    \includegraphics[width=0.48\textwidth]{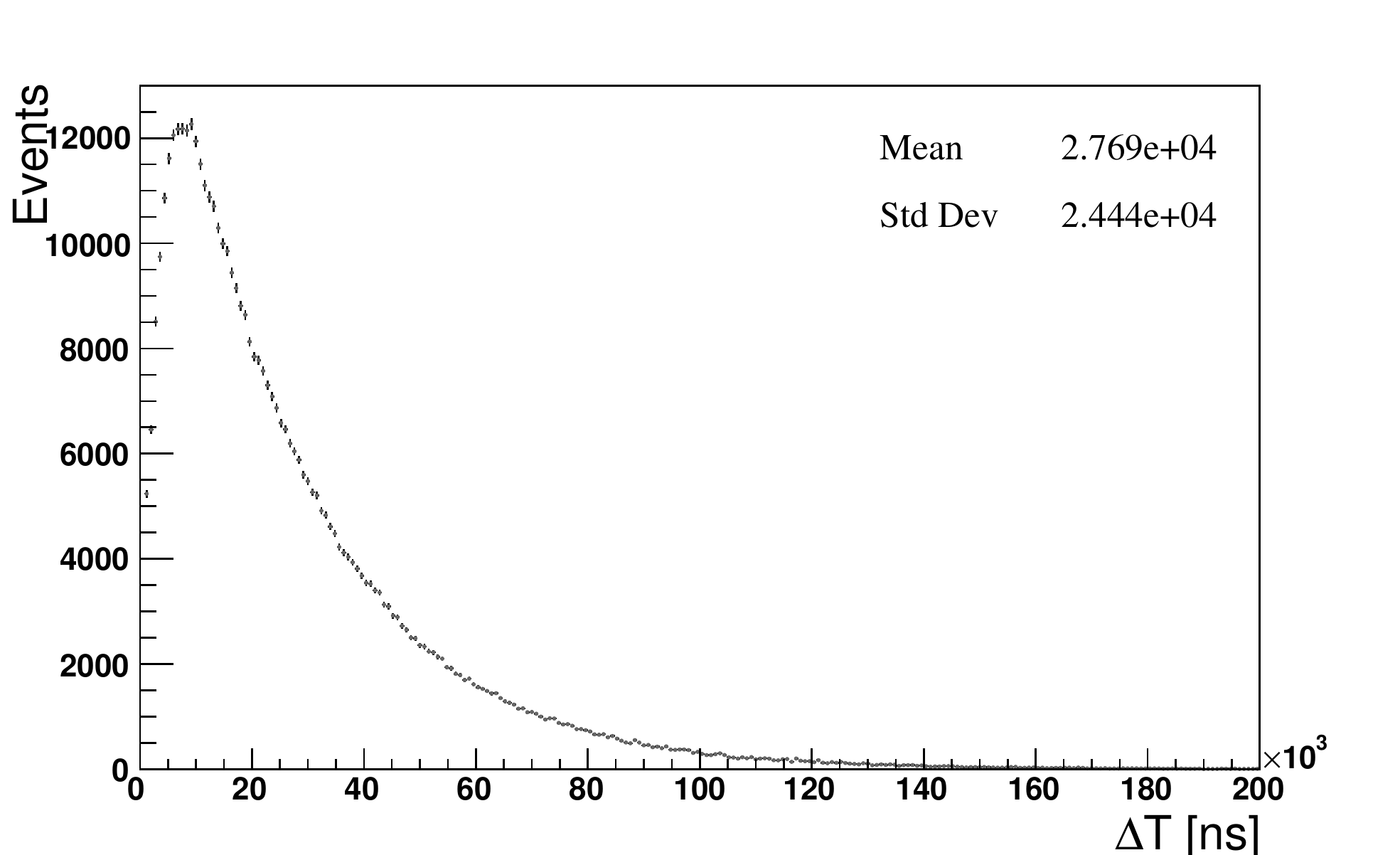}
    \caption{Time differences between prompt and delayed events for IBD events.}
    \label{fig:deltaT}
\end{figure}

The sum of those hits per event for both time ranges makes up the prompt and delayed energy distributions shown in Fig.~\ref{fig:prompt} and~\ref{fig:delayed}.
Both distributions have been scaled using the conversion factor obtained previously, from the gaussian fit in Fig.~\ref{fig:fit_MeV}, and expressed in term of visible energy. 
On Fig.~\ref{fig:delayed}, the ratio between the amplitudes of both peaks, at $2.2 \text{MeV}$ from neutron capture on hydrogen and at $\sim 8 \text{MeV}$ peak neutron captures on gadolinium, is directly dependent on the gadolinium concentration.
Higher gadolinium concentrations yield a higher number of captures in the latter peak and an increased detection efficiency above a fixed threshold.
As mentioned in section~\ref{sec:intro}, the time difference between the reconstructed prompt and delayed events is expected to behave like an exponential function with a constant, or rate, of $\sim$~30~$\mu$s. 
This behavior is confirmed in the simulation, as shown in Fig.~\ref{fig:deltaT}.

\section{Discussion}
\label{sec:5}
The detector design proposed and modeled in this study will be addressing the following motivations: 
\begin{itemize}
    \item Being the first neutrino detector built and operated in Turkey. In addition, it will act as an educational platform for neutrino researchers, provide hands-on experience for the development of new detector technologies, and train the next generation of scientists in Turkey. 
    \item Aiming to become a test-bed for experiments around the globe interested in testing and deploying new detector technologies. 
    \item Act as a near detector to the 30-ton detector mentioned in~\cite{FISCHER2020163931}. Operating both detectors would provide additional measurements of reactor-based neutrino oscillations at medium baseline ($>$1~km) and short  ($\leq$100~m) baseline. 
\end{itemize}

The simulation results obtained from this study show that a detector with a 3\% LS and 0.1\% Gd WbLS solution can adequately and efficiently detect neutrinos from the Akkuyu NPP. 
As mentioned previously, this particular medium was considered mainly because it has already been manufactured and could be ready to use on a short timescale. 
Since R\&D studies on WbLS are advancing quickly, thanks to a global effort and an interest from the particle and nuclear physics communities, one can imagine that more WbLS solutions with different liquid scintillator content will be manufactured on a large scale in the near future.
In anticipation of this possibility, simulation studies similar to the ones presented above were carried out with different WbLS mixtures.
Since the scintillation light yield increases with the WbLS liquid scintillator concentration, studies using a realistic 5\% LS mixture yield a better energy and vertex resolution.

Table~\ref{tab:mev} shows the energy and vertex resolutions for different WbLS mixtures. 

\begin{table}[!b]
    \centering
    \begin{tabular}{|c|c|c|}
    \hline
          Medium  &  Energy Resolution &Vertex Resolution \\
          \hline
          1\%LS     & $\sim 33 \%$    &  476  mm              \\
          3\%LS     & $\sim 24 \%$   &   406 mm                \\
          5\%LS     & $\sim 21 \%$    &  365  mm              \\
          \hline
    \end{tabular}
    \caption{The energy and vertex resolutions for different WbLS concentrations (1\%, \%3, and \%5) with a 0.1\% Gd.}
    \label{tab:mev}
\end{table}

This detector is intended to be used and operated as a testbed for experiments, such as THEIA~\cite{theia2020} and WATCHMAN~\cite{osti_1544490}, that expressed interest in using WbLS in addition to new photosensors in their detectors.
One of the main features that makes this combination of WbLS and new photosensors attractive for such experiments is the ability to separate Cherenkov and scintillation in the same event, thus providing a precise kinematic and calorimetric reconstruction of a detected particle.
While this separation was not explored in details in these simulation studies, it is the focus of several research groups both in terms of hardware and software.
While the effort on the software side focuses on machine learning based methods to select the fast Cherenkov component from the slow scintillation one, new hardware is being designed and developed to perform this task in-situ in the detector.
Dichroic filters are one way to separate these light sources at the hardware level.
With their ability to sort incoming light as a function of wavelengths, dichroic filters are an efficient, cheap and yet reliable way of separating the Cherenkov and scintillation components of the light emitted by a particle interacting in WbLS~\cite{Tanner2019, kaptanoglu2020spectral}. 
Such filters could be easily installed in the detector presented in his study, thanks to its design allowing for an easy access to its PMTs and inner structure.
Another method to separate the fast and the slow component of light is to use fast photosensors, able to separate all photon hits on a picosecond time scale, such as LAPPDs.
Such photodetectors can also be easily deployed in the detector, with only minimal changes to the inner structure \cite{back2017accelerator}.

While this study focused on the detector's design and its response to reactor neutrinos, one should note that deploying a detector in the vicinity of a nuclear reactor is a major challenge in terms of background mitigation.
If located more than 20~meters from a reactor core, direct backgrounds from fissions and fission products (gammas and neutrons) aren't a concern.
Similarly, accidental backgrounds from radioactive decays within the detector can be mitigated by carefully screening and choosing materials with a high level of radio-purity as well as utilizing the fact that, unlike IBD pairs, such events are not correlated in time.
As observed by the authors of~\cite{NUCIFER:2015hdd}, the main background we anticipate for such a detector is correlated background caused by cosmic muon interactions in, or in the vicinity of, the detector.
Such interactions are likely to cause spallation reactions that, in turn, are likely to emit neutrons with energies in the MeV range, or \textit{fast neutrons}.
With enough kinetic energy to generate a signal equivalent to a few MeV of visible energy followed by a capture on gadolinium, those neutrons can easily mimic an IBD pair.
Even assuming an efficient tagging of the muons entering the detector and its surroundings, this correlated background can not be fully excluded, especially with a detector placed at shallow depth such as the one modeled in this study.
Prior to a deployment, a dedicated background study must be carried out to understand this background and mitigate it using passive shielding and active vetos.

\section{Conclusion}
\label{sec:6}
Here, we present and discuss the simulation results of a 2-3 ton portable Gd-doped Water-based Liquid Scintillator (WbLS) detector as a Turkey's first low-energy neutrino detector near the Akkuyu Nuclear Power Plant (ANPP). 
The simulation studies in this paper are carried out using the RAT-PAC simulation package based on Geant4. 
The goal of this study was to study the response of a ton-scale detector to reactor neutrinos while investigating the feasibility of a deployment using existing mixtures of WbLS, with varying concentrations of liquid scintillator and gadolinium.
With an experimental setup consisting of a 1.5 m high and 1.5 wide cylindrical stainless steel tank instrumented by 24~PMTs, surrounded by scintillation paddles acting as an active cosmic veto and operated through basic high voltage and DAQ systems, this detector aims at being portable and easily deployed. 
This $\sim$2.5~ton detector filled with a 3\% LS and 0.1\% Gd WbLS solution would observe approximately 900 neutrino candidates per day if placed 50 meters away from the reactor cores.
Further studies will be carried out, in collaboration with the ANPP operators, to estimate the ideal location for this detector and assess the impact of correlated backgrounds at this location. 
This project is still in the proposal stage and, upon the approval of its funding, it will be built at the Detector R\&D Lab at Erciyes University for calibration and commissioning before being moved in the vicinity of the ANPP. 
This detector is intended to serve as a standalone detector able to monitor the thermal power of the ANPP in real-time, for nuclear non-proliferation R\&D purposes, but also a near detector for a planned program of short-baseline low energy neutrino studies in Turkey.
It will also be an opportunity to test new equipment and techniques for the growing field of neutrino physics, as well as train the next generation of neutrino and detector physicists under the Reactor Neutrino Experiments of Turkey (RNET) program. 

\begin{acknowledgements}
This work was supported by Scientific Research Projects (BAP) of Erciyes University, Turkey under the grant contract of FDS-2021-10856. It could not have been accomplished without the support and resources of Erciyes University. The authors owe special thanks to Watchman Collaboration for the Watchman version of the open-source Rat-Pac simulation program. The authors would also like to thank The Orebi Gann group at UC-Berkeley for developing and providing the WbLS cocktails in the Watchman GitHub.  
\end{acknowledgements}



\bibliographystyle{ieeetr}       
\bibliography{bibliography}   

%



\end{document}